\documentstyle[preprint,aps,version2]{revtex}  
\begin{document}
\thispagestyle{empty}

\renewcommand{\small}{\normalsize} 

\font\fortssbx=cmssbx10 scaled \magstep2
\hbox to \hsize{
\hskip.5in \raise.1in\hbox{\fortssbx University of Wisconsin - Madison}
\hfill\vtop{\hbox{\bf MAD/PH/878 }
            \hbox{UB-HET-95-02 }
            \hbox{March 1995}} }
\vspace{.25in}
 \begin{title}
{\bf Finite Width Effects and Gauge Invariance in \\
Radiative $W$ Production and Decay}
\end{title}
\author{U.~Baur$^1$ and \ D.~Zeppenfeld$^2$}
\begin{instit}
$^1$Department of Physics, State University of New York, Buffalo, NY
14260\\
$^2$Department of Physics, University of Wisconsin, Madison, WI 53706
\end{instit}
\begin{abstract}
\begin{center} {ABSTRACT} \end{center}
The naive implementation of finite width effects in processes involving
unstable particles can violate gauge invariance. For the example of
radiative $W$ production and decay,
$q\bar q' \to \ell\nu\gamma$, at tree level, it is demonstrated how
gauge invariance is restored by including the imaginary part of triangle
graphs in addition to resumming the imaginary contributions to the $W$
vacuum polarization. Monte Carlo results are presented for the Fermilab
Tevatron.
\end{abstract}

\newpage

Radiative $W$ boson production and decay at hadron colliders, {\it i.e.}
the process $q\bar q' \to \ell^\pm \nu\gamma$, is an important
testing ground for the Standard Model (SM). At large photon transverse
momentum this process allows the measurement of the $WW\gamma$ three gauge
boson coupling~\cite{Tev,BZ}. At small transverse momenta of the photon
or when the photon
is emitted collinearly to the final state charged lepton this
process needs to be fully understood when trying to extract precise
values of the $W$ boson mass and width from the Tevatron data: in the
CDF detector, for
example, photon emission effects in $W$ decays lead to a $W$ mass shift of
$-65\pm 20$~MeV ($-168\pm 20$~MeV) in the $W\to e\nu_e$
($W\to \mu\nu_\mu$) channel~\cite{Wmass}, and a shift in the $W$ width
of $70\pm 28$~MeV~\cite{Wwidth}.
However, initial state radiation diagrams are not included in the Monte
Carlo program currently used to simulate these effects~\cite{Wmass,Wwidth}.

Thus, a calculation is needed which fully describes initial
and final state radiation, which implements the effects of the lepton
mass, and which incorporates the finite $W$ decay width, $\Gamma_W$.
Finite lepton masses mostly affect
photon radiation in the collinear region. Finite $W$ width effects, on
the other hand, play an important role for large angle soft photon
bremsstrahlung, where
the radiative $W$ decay diagrams and $W\gamma$ production graphs may
interfere substantially. In Ref.~\cite{BK}, radiative $W$ decays were
considered, including a full treatment of lepton mass effects.
Initial state radiation was not incorporated into this
calculation. In Ref.~\cite{BZ}, we have calculated initial and final
state radiation effects. However, since we were interested in the
measurement
of the $WW\gamma$ coupling, our Monte Carlo program did not allow the
simulation of collinear final state photon radiation and the
implementation
of finite $W$ width effects was inadequate for soft photon emission.

Part of the problem can be traced to the difficulty of implementing finite
$W$ width effects while maintaining electromagnetic gauge
invariance~\cite{lopez}. The full
process $q\bar q' \to \ell^\pm \nu\gamma$ proceeds via Feynman graphs with
one
(initial and final state radiation) and two $W$ propagators (photon
emission
from the $W$). Replacing the $W$ propagator factors $1/(q^2-m_W^2)$ by a
Breit-Wigner form, $1/(q^2-m_W^2+im_W\Gamma_W)$, will disturb the gauge
cancellations between the individual Feynman graphs and thus lead to an
amplitude which is not electromagnetically gauge invariant when finite
lepton
mass effects are included. In addition, a constant imaginary part in the
inverse propagator is ad hoc: it results from fermion loop contributions
to the $W$ vacuum polarization and the imaginary part should vanish for
space-like momentum transfers.

In this Letter we demonstrate how this problem is solved by including the
imaginary part of $WW\gamma$ vertex corrections in addition to the
resummation of the $W$ vacuum polarization contributions. The resulting
amplitudes
respect electromagnetic gauge invariance, are valid for arbitrary
lepton masses, combine both initial and final state radiation,
exhibit good high energy behaviour, and the imaginary parts of the
propagators are exactly as derived from the $W$ vacuum polarization.
Even though we are only considering radiative $W$ production and decay
here, our
method can immediately be generalized to $t$-channel processes such as
$e^+e^-\to e^+W^-\nu_e\to e^+\nu_e d\bar u$~\cite{vO,shimizu}. We also
present some numerical results of our Monte Carlo calculation.

The general structure is best understood by first considering the lower
order process ($q\bar q' \to \ell^- \bar\nu$ to be specific) without
photon emission. The corresponding Feynman graphs are shown in
Fig.~\ref{figW}. Finite width effects are included in a tree level
calculation by resumming the imaginary part of the $W$ vacuum polarization.
When neglecting the fermion masses in the
loops of Fig.~\ref{figW} the transverse part of the $W$ vacuum
polarization
receives an imaginary contribution
\begin{equation}\label{Pi_W^T}
Im\, \Pi_W^T(q^2) = \sum_f {g^2\over 48\pi} q^2
= q^2 {\Gamma_W\over m_W}\; ,
\end{equation}
while the imaginary part of the longitudinal piece vanishes. In the
unitary
gauge the $W$ propagator is thus given by
\begin{eqnarray}
D_W^{\mu\nu}(q) & = &{-i \over q^2-m_W^2 + i Im\, \Pi_W^T(q^2)}
\left( g^{\mu\nu}-{q^\mu q^\nu \over q^2}\right)
+{i\over m_W^2 - i Im\, \Pi_W^L(q^2)}\; {q^\mu q^\nu \over q^2}
\nonumber \\[3.mm]
& = & {-i \over q^2-m_W^2 + i q^2 \gamma_W}
\left( g^{\mu\nu}-{q^\mu q^\nu \over m_W^2}
(1+ i \gamma_W )   \right)\; ,  \label{Wprop}
\end{eqnarray}
where we have used the abbreviation $\gamma_W = \Gamma_W/m_W$. Note that
the $W$ propagator has received a $q^2$ dependent effective width which
actually
would vanish in the space-like region. Since here we are only interested
in (virtual) $W$ decays we do not include the corresponding step-function
factor.

A gauge invariant expression for the amplitude of the process
$q\bar q' \to \ell^- \bar\nu\gamma$ is obtained by attaching the final
state photon in all possible ways to all charged particle propagators
in the Feynman graphs of Fig.~\ref{figW}. This includes radiation off the
two incoming quark lines, radiation off the final state charged lepton,
and radiation off the $W$ propagators. In addition, the photon must be
attached to the charged fermions inside the $W$ vacuum polarization
loops, leading to the fermion triangle graphs of Fig.~\ref{figvertex}.
Since we are only keeping the imaginary part of $\Pi_W^T(q^2)$,
consistency
requires including the imaginary part of the
triangle graphs only. This imaginary part is obtained by cutting the
triangle graphs into on-shell intermediate states in all possible ways,
as shown in the figure.

The full set of Feynman graphs contributing to $q\bar
q'\to\ell^- \bar\nu\gamma$ is depicted in Fig.~\ref{figWgamma}.
When attaching the photon in all possible ways to either one of the
fermion
loops or to one of the lowest order $W$ propagators in all graphs of
Fig.~\ref{figW}, the remaining sum over $W$ vacuum polarization graphs
restores the full $W$ propagator of Eq.~(\ref{Wprop}) on either side of
the triangle graph or the $WW\gamma$ vertex. Hence, after resummation, one
obtains the Feynman graph of Fig.~\ref{figWgamma}c where the full
$WW\gamma$
vertex is given by the sum of the lowest order vertex and the imaginary
part of the triangle graphs, as defined in Fig.~\ref{figvertex}.

For the momentum flow shown in Fig.~\ref{figvertex} the lowest order
vertex is given by the familiar expression
\begin{equation}\label{LOWWgvertex}
-ie\Gamma_0^{\alpha\beta\mu} = -i e \left( (q_1+q_2)^\mu g^{\alpha\beta}
                            - (q_1+k)^\beta g^{\mu\alpha}
                            + (k-q_2)^\alpha g^{\mu\beta} \right)\; .
\end{equation}
Neglecting the masses of the fermions in the triangle graphs and dropping
terms proportional to $k^\mu$ (which will be contracted with the photon
polarization vector $\varepsilon^{*\mu}$ and hence vanish in the
amplitude)
the contributions from the four triangle graphs reduce to an extremely
simple form. Each fermion doublet $f$, irrespective of its hypercharge,
adds $i(g^2/48\pi)\Gamma_0$ to the lowest order $WW\gamma$ vertex
$\Gamma_0$.
After summing over all fermion species, the lowest order vertex is thus
replaced by
\begin{equation}\label{WWgvertex}
\Gamma^{\alpha\beta\mu} =
\Gamma_0^{\alpha\beta\mu}\left(1+\sum_f {ig^2\over 48 \pi}\right) =
\Gamma_0^{\alpha\beta\mu}\left(1+i\,{\Gamma_W\over m_W}\right) =
\Gamma_0^{\alpha\beta\mu}\left(1+i\gamma_W \right)\; .
\end{equation}
This expression corresponds to the full $WW\gamma$ vertex which is
represented
by the hatched blobs in Figs.~\ref{figvertex} and~\ref{figWgamma}.

By construction, the resulting amplitude for the process
$q\bar q' \to \ell^- \bar\nu  \gamma $ should be gauge invariant.
Indeed, gauge invariance of the full amplitude can be traced to the
electromagnetic Ward identity~\cite{lopez}
\begin{equation}\label{ward}
k_\mu {\Gamma_{\alpha\beta}}^\mu = \left(
iD_W\right)^{-1}_{\alpha\beta}(q_1)
- \left( iD_W\right)^{-1}_{\alpha\beta}(q_2)\; .
\end{equation}
Since
\begin{equation}\label{ward1}
k_\mu \Gamma^{\alpha\beta\mu} = \left(
( q_1^2 g^{\alpha\beta} - q_1^\alpha q_1^\beta ) -
( q_2^2 g^{\alpha\beta} - q_2^\alpha q_2^\beta ) \right)
\left(1+i\gamma_W\right)\;,
\end{equation}
and
\begin{equation}\label{ward2}
\left( iD_W\right)^{-1}_{\alpha\beta}(q) =
\left( q^2-m_W^2+iq^2\gamma_W \right)
\left( g_{\alpha\beta} - {q_\alpha q_\beta \over q^2} \right)
-m_W^2 {q_\alpha q_\beta \over q^2}\; ,
\end{equation}
this Ward identity is satisfied for the $W$ propagator and $WW\gamma$
vertex of Eqs.~(\ref{Wprop}) and (\ref{WWgvertex}).

There are other ways to include finite width effects while restoring
electromagnetic gauge invariance~\cite{vO}. One possibility is to multiply
the complete tree level amplitude by an overall factor
$F(q^2)= (q^2-m_W^2)/(q^2-m_W^2+im_W\Gamma_W)$ for each almost on-shell
$W$ propagator~\cite{BVZ}. This procedure implements Breit-Wigner
propagators in all resonant graphs but multiplies non-resonant
contributions by an ad-hoc factor. This is phenomenologically
acceptable if the non-resonant contributions are sub-leading.
For $W\gamma$ production, however, the ``non-resonant'' graphs
of Fig.~\ref{figWgamma}a,b are dominant near threshold and multiplying
them by the factor $F(\hat s)$ decreases {\it e.g.} the soft photon
emission cross section by 30\%
or more as compared to the correct treatment described above.

Electromagnetic gauge invariance is also obtained by dropping the
triangle graph contributions and using a naive Breit Wigner
propagator~\cite{lopez},
\begin{equation}
D_{W,{\rm naive}}^{\mu\nu}(q) = {-i \over q^2-m_W^2 + i m_W \Gamma_W}
\left( g^{\mu\nu}-{q^\mu q^\nu \over m_W^2-im_W\Gamma_W} \right)\; .
\label{Wpropnaive}
\end{equation}
However, this propagator does not correspond to a resummation of
$W$ vacuum polarization graphs, the imaginary part does not vanish in the
space-like region, and the imaginary part of the $W$ mass squared in the
longitudinal piece of the propagator is entirely unphysical.
On theoretical grounds this solution of the
gauge-invariance problem must therefore be rejected, even though, for the
problem at hand, numerical results are very similar to the full treatment
including triangle graphs.

The modification of the lowest order $WW\gamma$ vertex in
Eq.~(\ref{WWgvertex})
looks like the introduction of anomalous couplings $g_1^\gamma =
\kappa_\gamma
= 1+i\gamma_W$ and one may thus worry that the full amplitude will violate
unitarity at large center of mass energies $\sqrt{\hat s}$. While indeed
the
vertex is modified, this modification is compensated by the effective
$\hat s$-dependent width in the propagator. As compared to the
expressions with a lowest order propagator, $1/(\hat s - m_W^2)$,  which
of course has good high energy behaviour, the overall effect is
multiplication of the analog of the Feynman graph of Fig.~\ref{figWgamma}c
by a factor
\begin{equation}
G(\hat s) = {\hat s - m_W^2 \over \hat s(1+i\gamma_W) - m_W^2}\;
(1+i\gamma_W) = 1\; -\;
{i\Gamma_W m_W \over \hat s - m_W^2 + im_W\Gamma_W{\hat s\over m_W^2}}\; .
\label{formfactor}
\end{equation}
Obviously, $G(\hat s)\to 1$ as $\hat s \to \infty$ and the high energy
behaviour of our finite width amplitude is identical to the one of the
naive
tree level result for $W\gamma$ production. In fact, the contributions
from the triangle graphs are crucial to compensate the bad high energy
behaviour introduced by the $q^2$-dependent width in Eq.~(\ref{Wprop}).

What are the results of our implementation of finite width, fermion
triangle, and lepton mass effects? In Table~I we give the fraction of
$W^\pm \to \ell^\pm \nu$ events ($\ell=e,\,\mu$) at the Tevatron
($p\bar p$ collisions at $\sqrt{s}=1.8$~TeV) which contain a photon
with a transverse
energy $E_T(\gamma)>E_T^0(\gamma)$. We then compare our results
with the expectation for final state radiation only
(in the narrow $W$-width approximation), as in Ref.~\cite{BK}.
Specifically, we consider the phase space region
\begin{eqnarray}\label{cuts1}
p_T(\ell)>25~{\rm GeV}\;, \qquad p\llap/_T > 25~&{\rm GeV}&\;, \qquad
|\eta(\ell)|<1\; , \qquad |\eta(\gamma)|<3.6 \nonumber \\   
65~{\rm GeV}< & m_T^W &< 100~{\rm GeV}\; ,
\end{eqnarray}
where $m_T^W$ denotes the transverse mass of the $\ell\nu$ system. The
phase space region defined by Eq.~(\ref{cuts1}) is
very similar to that used by CDF~\cite{Wmass} and D\O~\cite{dzero} to
extract the $W$ mass.
We then vary the requirements on the minimum photon transverse energy,
$E_T^0(\gamma)$, and the lepton photon separation, $\Delta
R(\ell,\gamma)$.
QCD corrections and structure function effects largely cancel in the
cross section ratio
considered. Finite lepton masses are seen to hardly influence large
angle photon radiation whereas the inclusive rate for photon radiation
in the electron channel is about a factor two larger than the result
obtained for muons. Initial state photon radiation approximately triples
the fraction of $W$ boson events which contain a well isolated photon
($\Delta R(\ell,\gamma)>0.7$).
It increases the total number of $W\to e\nu$ ($W\to\mu\nu$) events which
contain a photon by typically $\sim 10\%$ ($\sim 20\%$).
The event fractions presented in Table~I are insensitive to the maximum
absolute lepton pseudorapidity, $\eta_{\rm max}$,
in the range $\eta_{\rm max}=0.6\dots 1.2$.

Finally, we briefly comment on radiative $Z$ production and decay,
$q\bar q\to Z\gamma\to\ell^+\ell^-\gamma$ and $q\bar q\to
Z\to\ell^+\ell^-\gamma$. Since only diagrams with one $Z$ propagator
contribute, the gauge invariance problems encountered in the $W$ case
when incorporating finite lepton masses do not arise here, and the
calculation is straightforward. Using the results of Ref.~\cite{babe},
we have calculated radiative $Z$ production and decay including finite
lepton masses at hadron colliders. The fraction of $Z$ boson events which
contains a photon with
$E_T(\gamma)>E_T^0(\gamma)$ is about a factor~2 larger than the
corresponding number in the $W$ case. For $\Delta R(\ell,\gamma)>0.7$,
initial state radiation approximately doubles the number of $Z$ boson
events with a photon.

\acknowledgements
We would like to thank M.~Demarteau, S.~Errede and Y.-K.~Kim for many
stimulating discussions.
This research was supported in part by the University of Wisconsin
Research Committee with funds granted by the Wisconsin Alumni Research
Foundation and in part by the U.~S.~Department of Energy under Grant
No.~DE-FG02-95ER40896.

\newpage             

\newpage
\begin{table}
\caption{Fraction of $W\to e\nu$ and $W\to\mu\nu$ events (in percent)
at the Tevatron ($p\bar p$ collisions at $\sqrt{s}=1.8$~TeV).
We consider events in the phase space region of Eq.~(\ref{cuts1})
which contain a photon with
transverse energy $E_T(\gamma)>E_T^0(\gamma)$ for two different minimal
separations between the charged lepton and the photon, $\Delta
R(\ell,\gamma)$. The results obtained by considering
final state radiation only (in the narrow $W$ width approximation) are
shown in brackets. \\[3.mm]}
\begin{tabular}{ccccc}
\multicolumn{1}{c}{} & \multicolumn{2}{c}{$W\to e\nu$} &
\multicolumn{2}{c}{$W\to\mu\nu$} \\
$E_T^0(\gamma)$ & $\Delta R(e,\gamma)>0$ & $\Delta R(e,\gamma)>0.7$ &
$\Delta R(\mu,\gamma)>0$ &$\Delta R(\mu,\gamma)>0.7$ \\
\tableline
0.1 & 19.3 (17.5) & 3.63 (1.31) & 11.8 (9.8) & 3.62 (1.32) \\
0.3 & 14.8 (13.5) & 2.56 (0.95) & 8.76 (7.37) & 2.54 (0.95) \\
1 & 9.42 (8.63) & 1.45 (0.57) & 5.37 (4.62) & 1.44 (0.56) \\
3 & 4.58 (4.14) & 0.66 (0.24) & 2.55 (2.14) & 0.65 (0.24) \\
10 & 0.55 (0.36) & 0.13 (0.01) & 0.32 (0.17) & 0.13 (0.01) \\
30 & 0.014 (--) & 0.012 (--) & 0.013 (--) & 0.012 (--) \\
\end{tabular}
\end{table}

\newpage

\figure{
\label{figW}
Feynman graphs for the lowest order process $q\bar q'\to \ell^-\bar\nu$.
The resummation of the imaginary part of the $W$ vacuum polarization
leads to the Breit-Wigner type $W$ propagator of Eq.~(\ref{Wprop}) which
is represented by the shaded blob.
}

\figure{
\label{figvertex}
Effective $WW\gamma$ vertex as needed in the tree level calculation of
$W\gamma$ production. In addition to the lowest order vertex the imaginary
parts of the fermion triangles must be included (see
Eq.~(\ref{WWgvertex})).
}

\figure{
\label{figWgamma}
Feynman graphs for the process $q\bar q' \to \ell^- \bar\nu\gamma$. The
shaded and hatched blobs indicate the full $W$ propagators as defined
in Fig.~\ref{figW} and the full $WW\gamma$ vertex as given in
Fig.~\ref{figvertex}, respectively.
}

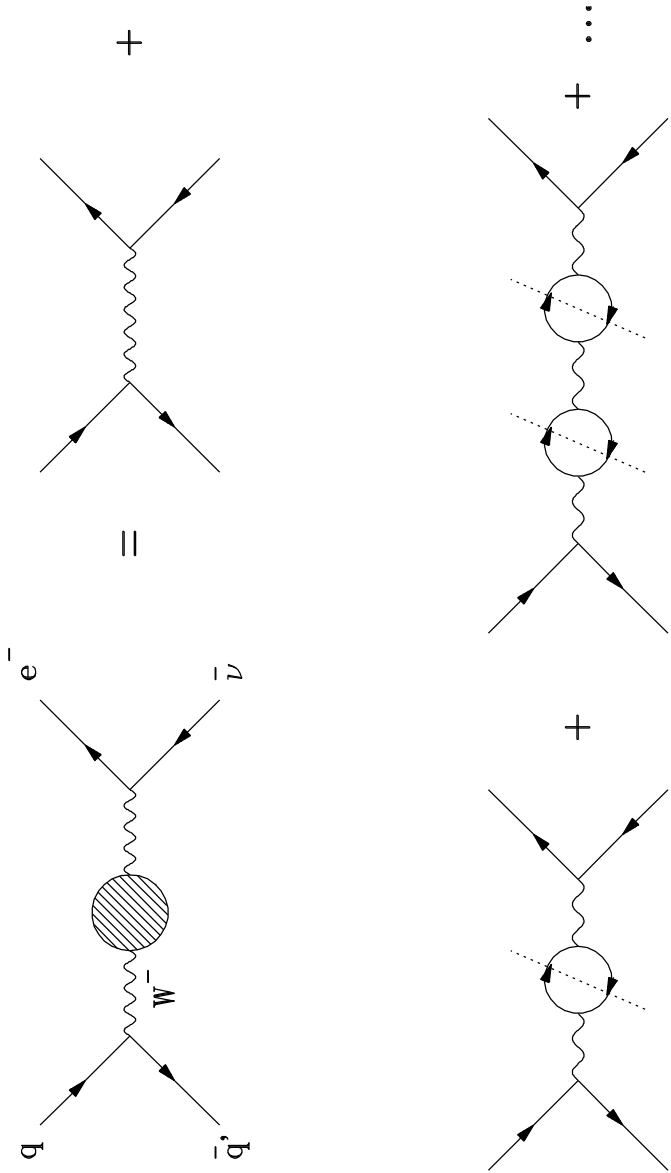
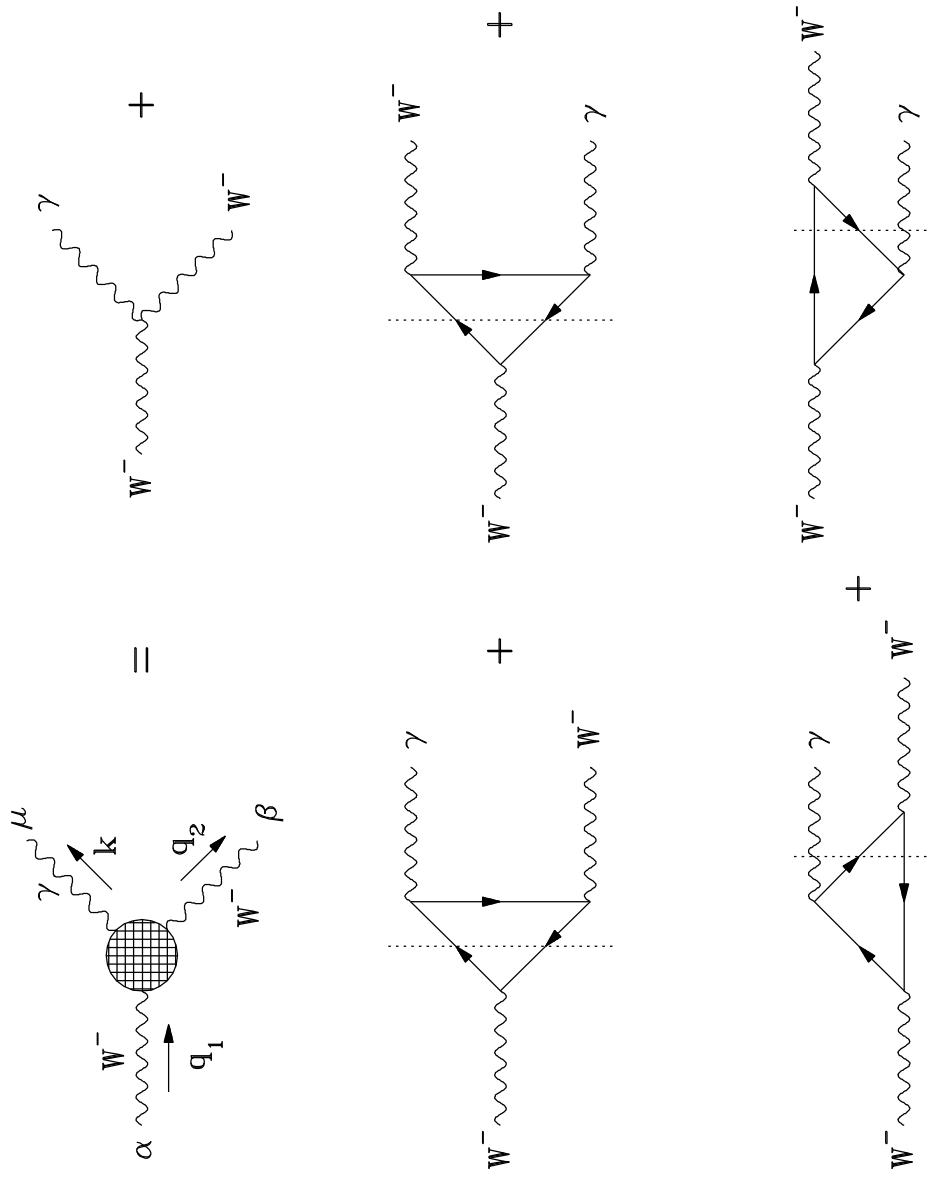
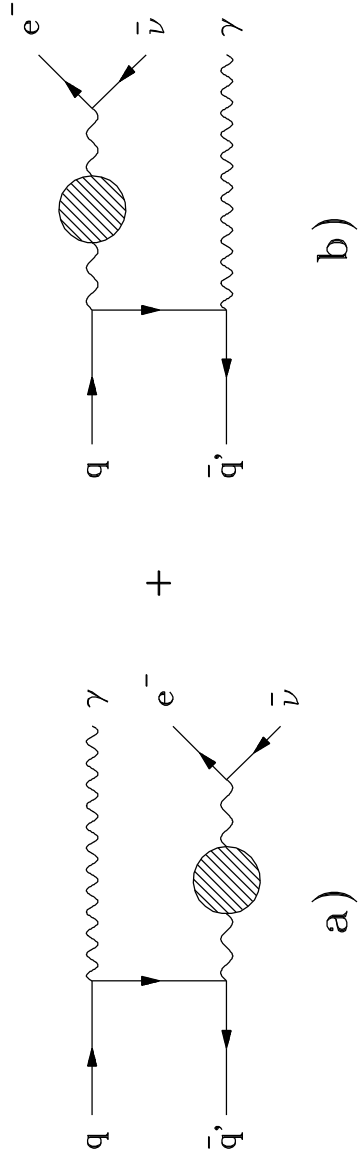
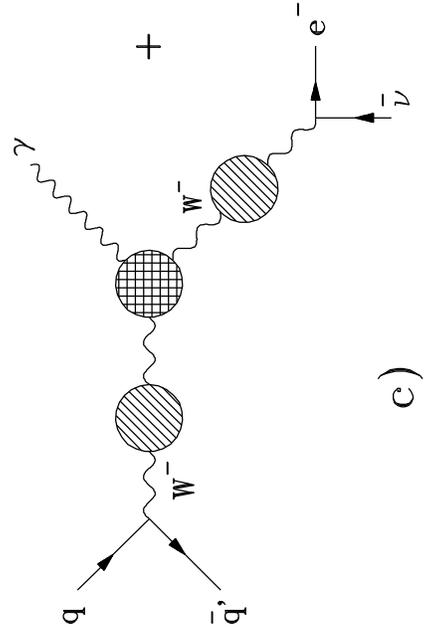
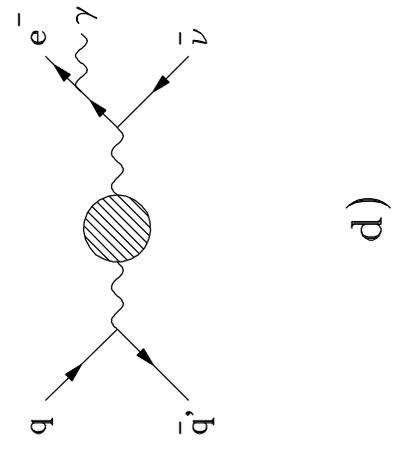
\end{document}